%% file: ESSS_2015.tex
\documentclass[copyright,creativecommons,noderivs,noncommercial,a4paper]{eptcs}

\usepackage{amssymb}
\usepackage{caption}
\usepackage{subfigure}
\usepackage{multirow}
\usepackage{paralist}
\usepackage{algorithm2e}
\usepackage{amsmath}
\usepackage{hhline}
\usepackage{verbatim}
\usepackage{program}
\usepackage{graphicx}
\usepackage[table]{xcolor}
\usepackage{url}
\usepackage{makeidx}
\usepackage{graphicx}
\usepackage{epstopdf} 
\usepackage{wrapfig}

\usepackage{color}
\usepackage{listings}
\lstdefinelanguage{pvs}
{morekeywords={THEORY,FORALL,THEOREM,EXISTS,AND,IF,ELSE,THEN,ENDIF,TABLE,ENDTABLE,COND,ENDCOND,IMPLIES,LEMMA,LET,IN,RECURSIVE,MEASURE,DO,TO,BY,FOR,END_FOR,VAR_INPUT,VAR_OUTPUT,VAR,END_VAR,TYPE,TRUE,FALSE,IMPORTING,END,PRED,BOOL,LAMBDA,NOT},
sensitive=false,
morecomment=[l]{\%},
morecomment=[s]{/*}{*/},
morestring=[b]",
}
\lstnewenvironment{pvs}[1][]
    {\lstset{#1,captionpos=b,language=pvs,
    mathescape=true,
    basicstyle=\small\ttfamily,
    numbers=none,
	frame=single,
    firstnumber=auto
    }}
    {}

\urldef{\mailsa}\path|{alfred.hofmann, ursula.barth, ingrid.haas, frank.holzwarth,|
\urldef{\mailsb}\path|anna.kramer, leonie.kunz, christine.reiss, nicole.sator,|
\urldef{\mailsc}\path|erika.siebert-cole, peter.strasser, lncs}@springer.com|

\usepackage{fancyvrb, courier} 

\DefineVerbatimEnvironment{code}{Verbatim}{frame=none, framesep=10mm, fontsize=\small, commandchars=\\\{\}}

\newcommand{\bigbrace}[1]{
    \left( 
        \begin{array}{l}
        #1
        \end{array} 
    \right)
}

\usepackage{wrapfig}

\usepackage{hyperref}
\hypersetup{%
  colorlinks=false,
  linkbordercolor=red,
  pdfborderstyle={/S/U/W 1}
}

\newcommand\name[1]{$\textit{#1}$}

\makeatletter
 \newcommand\npar{\@startsection{subsection}{2}{\z@}{-2\p@ \@plus -4\p@ \@minus -4\p@}{-0.5em \@plus -0.22em \@minus -0.1em}{\normalfont\large\bfseries}}
\makeatother

\title{Formal Verification of Real-Time Function Blocks Using PVS}

\author{
Linna Pang, Chen-Wei Wang, Mark Lawford, Alan Wassyng
\institute{McMaster Centre for Software Certification, McMaster University, Canada L8S 4K1}
\email{\{pangl,wangcw,lawford,wassyng\}@mcmaster.ca}
\and
Josh Newell, Vera Chow, and David Tremaine
\institute{Systemware Innovation Corporation, Toronto, Canada M4P 1E4}
\email{\{jnewell,vchow,tremaine\}@swi.com}}

\def\authorrunning{Linna Pang et al.}
\def\titlerunning{Formal Verification of Real-Time Function Blocks Using PVS}

\begin{document}

\maketitle

\begin{abstract}

A critical step towards certifying safety-critical systems is to check their conformance to hard real-time requirements. A promising way to achieve this is by building the systems from pre-verified components and verifying their correctness in a compositional manner. We previously reported a formal approach to verifying function blocks (FBs) using tabular expressions and the PVS proof assistant. By applying our approach to the IEC~61131-3 standard of Programmable Logic Controllers (PLCs), we constructed a repository of precise specification and reusable (proven) theorems of feasibility and correctness for FBs. However, we previously did not apply our approach to verify FBs against timing requirements, since IEC~61131-3 does not define composite FBs built from timers. In this paper, based on our experience in the nuclear domain, we conduct two realistic case studies, consisting of the software requirements and the proposed FB implementations for two subsystems of an industrial control system. The implementations are built from IEC~61131-3 FBs, including the on-delay timer. We find issues during the verification process and suggest solutions. 

\end{abstract}

\input{input_sec_introduction.tex}

\input{input_sec_preliminaries.tex}

\input{input_sec_ton_with_tolerances.tex} 

\input{input_sec_case_study_trip_sealin.tex}

\input{input_sec_case_study_pushbutton.tex}

\input{input_sec_lessons.tex}

\input{input_sec_related_works.tex}
\input{input_sec_conclusion.tex}


\bibliographystyle{eptcs}
\bibliography{Linna_ResearchProposal}

\end{document}

%% file: input_sec_introduction.tex
\section{Introduction}\label{sec:intro}

Many industrial safety-critical software control systems are based upon Programmable Logic Controllers (PLCs). Function blocks (FBs) are reusable components for implementing the behaviour of PLCs in a hierarchical way. In one of its supplements, the aviation standard DO-178C~\cite{DO178C} advocates the use of formal methods to construct, develop, and reason about mathematical models of system behaviours. Applying the principles of DO-178C to PLC-based systems, we may obtain high-quality PLCs by: 1) pre-verifying standard FBs using formal methods; 2) building systems from pre-verified components; and 3) verifying their correctness in a compositional manner.

We recently reported a formal methodology~\cite{PangWangLawfordWassyng2013} for specifying requirements for FBs, and for verifying the correctness of their implementations expressed in, e.g., function block diagrams (FBDs). In our approach, we use tabular expressions (a.k.a. function tables)~\cite{Parnas:1994:PDW:203102.203107} for specification and the PVS proof assistant~\cite{Owre1992} for formal verification. Tabular expressions are a way to document system requirements as black-box, input-output relations that has proven to be practical and effective in industry~\cite{WasLaw:2003}. PVS provides an integrated environment with mechanized support for writing specifications using tabular expressions and (higher-order) predicates, and for (interactively) proving that implementations satisfy the tabular requirements using sequent-style deductions. We successfully applied our approach to the FB library of IEC~61131-3~\cite[Annex F]{IEC:2003:IEP}, an industrial standard for PLCs, resulting in a repository of: 1) precise specifications of input-output requirements; and 2) reusable theorems of feasibility and correctness for the FB library. 

A critical step towards certifying safety-critical systems is to check their conformance to hard real-time requirements. An \emph{implementable} timing requirement must specify \emph{tolerances} to account for various factors --- e.g., sampling rates, computation time, and latency --- that will delay the software controller's response to its operating environment. A common type of functional timing requirements specifies that a monitored condition \name{C} must sustain over a time duration, say \name{timeout}, with tolerances $-\delta L$ and $+\delta R$, before being detected by the controller. Such sustained timing requirements may be formalized using an infix \name{Held\_For} operator~\cite{Wassyng05}. For example, we write 
\begin{equation}
(~(\textit{signal} \geq \textit{setpoint}) \; \textit{Held\_For}\; (300,\ -50,\ +50)~) \Rightarrow (\textit{c\_var} = \textit{trip})\label{eq:held_for}
\end{equation}
to specify that a sensor signal going out of its safety range should cause a ``trip'' if it sustains for longer than 350 ms, and should not if it lasts for less than 250 ms (to filter out the effect of a noisy signal). 

The requirement specified in Eq.~\ref{eq:held_for} is \textit{non-deterministic} since it allows any implementation that trips when $\textit{signal} \geq \textit{setpoint}$ sustains between $[250ms,\ 350ms]$. As we will see in our two case studies, such a simple requirement can be used as part of specifying more complex real-time behaviour. To resolve such non-determinism, at the requirements level we adopt a deterministic operator \name{Held\_For\_I}~\cite[p. 86]{Hu08}, which becomes \name{true} at the first sampling point after the monitored condition has been enabled for $\textit{d} - \delta \textit{L}$ time units. For example, by substituting the expression $((\textit{signal} \geq \textit{setpoint})\; \textit{Held\_For\_I}\; (300 - 50))$ into Eq.~\ref{eq:held_for}, we specify that the triggering condition sustaining for longer than 250 ms should cause a ``trip''. Similarly, at the implementation level we adopt the \name{Timer\_I} operator~\cite[p. 98]{Hu08} for counting the elapsed time of some monitored condition. The relationship between these two operators, at levels of requirements and implementation, is proved as a theorem \name{TimerGeneral\_I}~\cite[p. 99]{Hu08}: $(\textit{C}\; \textit{Held\_For\_I}\; (timeout - \delta \textit{L}))$ is equivalent to $(\textit{Timer\_I}~(C) \geq \textit{timeout} - \delta \textit{L})$. See also Sec.~\ref{sec:preliminary}.

We previously did not apply our approach~\cite{PangWangLawfordWassyng2013} to verify FBs against this type of more complex timing requirements because IEC~61131-3 only includes simple timer blocks (i.e, on-delay, off-delay, and pulse timers), but not any more complex FBs built from those timers. Furthermore, our requirements model for IEC~61131-3 timers~\cite{PangWangLawfordWassyng2013} describes idealized behaviour: as the monitored condition becomes enabled, the timer instantaneously responds (i.e., starts counting the duration of enablement), not considering sampling, computational delays, and timing tolerances. 

Based on our experience on the Darlington Nuclear Shutdown Systems Trip Computer Software Redesign Project~\cite{WasLaw:2003}, and motivated by anticipated FB based projects, we address the above issues by conducting realistic case studies. Each case study consists of the software requirements and FBD implementation for a subsystem of an industrial control system. Implementations are built from IEC~61131-3 FBs, including the on-delay timer to implement more complex real-time behaviour. 

\begin{figure}
\begin{center}
\includegraphics[width=\textwidth]{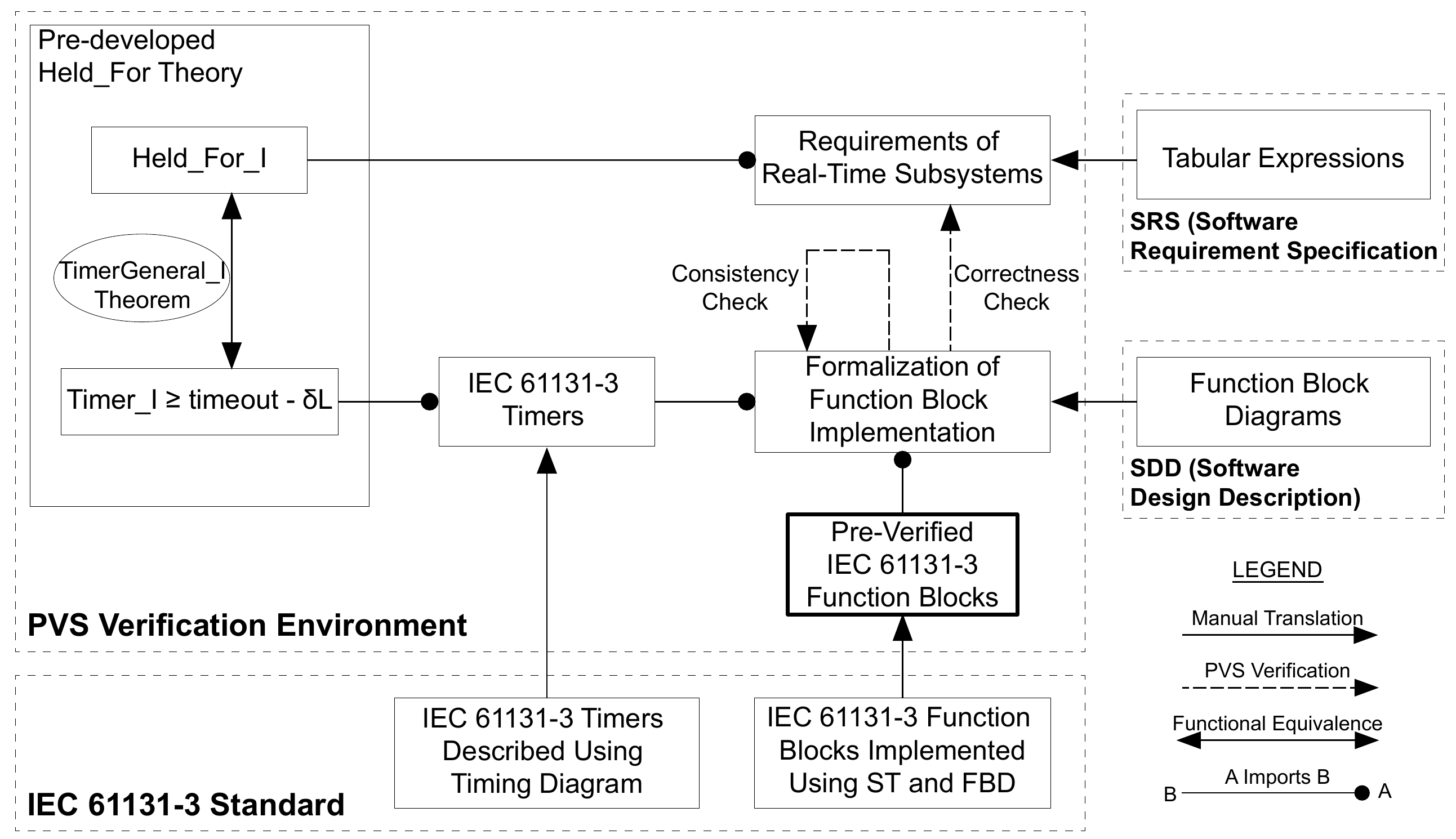}
\caption{Framework for verification of FB based systems timing requirements}\label{fig:Framework}
\end{center}
\end{figure}

Fig.~\ref{fig:Framework} summarizes our verification process and contributions. To incorporate the notion of tolerances, we reuse the timing operators \name{Held\_For\_I} (to formalize requirements) and \name{Timer\_I} (to formalize implementations) from \cite{Hu08}. The verification goal is that the proposed FBD implementations, included in the Software Design Description (SDD), are: (a) \emph{consistent}, or feasible, meaning that an output can always be produced on valid inputs; and (b) \emph{correct} with respect to the timing requirements specified using \name{Held\_For\_I}, included in the Software Requirements Specification (SRS). This work builds on our previous results of verifying IEC~61131-3 FBs~\cite{PangWangLawfordWassyng2013} that provide a sound semantic foundation for formalizing and verifying PLC programs expressed using FBDs. 

There are four contributions of this paper. First, to incorporate tolerances, we use the \name{Timer\_I} operator to re-formalize all three IEC~61131-3 timer (Sec.~\ref{sec:timers}). Second, for the representative subsystems we study (one with a feedback loop presented in Sec.~\ref{sec:Sealin} and the other in Sec.~\ref{sec:PB}), we use the re-formalized IEC~61131 timers for their proposed FBD implementations, and prove that they are feasible and satisfy the intended timing requirements in SRS.  Third, we find issues of initialization failure (Sec.~\ref{sec:Sealin}) and missing implementation assumptions (Sec.~\ref{sec:PB}), and suggest possible solutions. Fourth, we identify patterns of proof commands (Sec.~\ref{sec:lessons}) that are amenable to strategies (or proof scripts) that will facilitate the automated verification of the feasibility and correctness of other subsystems. 


\smallskip

\noindent\textit{Resources}. Sources of the case studies (verified using PVS~6.0) are available at \url{http://www.cas.mcmaster.ca/~lawford/papers/ESSS2015}. Background theories (e.g., \name{Held\_For\_I}, \name{Timer\_I}, etc.) and complete details of case studies covered in this paper are included in an extended report~\cite{PangWangLawfordWassyngTechReport2014}.




%% file: input_sec_preliminaries.tex
\section{Preliminaries}\label{sec:preliminary}

We review the use of tabular expressions, the relevant PVS theories of timing operators at levels of requirements and implementation, and the formal verification approach~\cite{PangWangLawfordWassyng2013} that is adapted in our two case studies (Sec.~\ref{sec:Sealin} and Sec.~\ref{sec:PB}). 

\smallskip

\npar{Tabular Expressions}\label{sec:preliminaries:tables}

Tabular expressions (a.k.a. function tables)~\cite{Parnas:1994:PDW:203102.203107} are an effective approach for describing conditionals and relations, thus ideal for documenting many system requirements. They are arguably easier to comprehend and to maintain than conventional mathematical expressions. Tabular expressions have well-defined formal semantics (e.\,g.,~\cite{Jin2010980}). For our purpose of capturing the input-output requirements of timing function blocks, the tabular structure in Fig.~\ref{fig:HCT} below suffices: rows in the first column denote input conditions, and rows in the second column denote the corresponding output results. The input column may be sub-divided to specify sub-conditions. When the output column denotes a state variable, we may write \name{NC} to abbreviate the case of ``no change'' on its value.

\begin{figure}[htb]
\begin{center}
\newcommand{\s}[2]{$#1_{#2}$}
\begin{minipage}{\textwidth}\centering
\begin{minipage}{.25\textwidth}
{\small \begin{tabular}{|c|cc}
\multicolumn{2}{c}{}
&\multicolumn{1}{c}{\small{\textit{Result}}}\\
\hhline{~~-}
\multicolumn{2}{c}{\small{\textit{Condition}}}& \multicolumn{1}{||c||}{\cellcolor[gray]{0.8}\bf{f}}\\ \hline
\multirow{4}{*}{\s{C}{1}} & \s{C}{1.1} & \multicolumn{1}{||c||}{\s{res}{1}} \\ \cline{2-3}
                          & \s{C}{1.2} & \multicolumn{1}{||c||}{\s{res}{2}} \\ \cline{2-3}
                          & $\dots$ & \multicolumn{1}{||c||}{$\dots$} \\ \cline{2-3}
                          & \s{C}{1.m} & \multicolumn{1}{||c||}{\s{res}{m}} \\ \cline{1-3}
 \multicolumn{2}{|c}{$\dots$} & \multicolumn{1}{||c||}{$\dots$} \\ \hline
\multicolumn{2}{|c}{\multirow{1}{*}{\s{C}{n}}} & \multicolumn{1}{||c||}{\s{res}{n}}\\\hline
\end{tabular}
}\end{minipage}
\begin{minipage}{.4\textwidth}   
\begin{code}
\textbf{IF} \s{C}{1}
  \textbf{IF}     \s{C}{1.1} \textbf{THEN} f = \s{res}{1}
  \textbf{ELSEIF} \s{C}{1.2} \textbf{THEN} f = \s{res}{2}
  ...
  \textbf{ELSEIF} \s{C}{1.m} \textbf{THEN} f = \s{res}{m}
\textbf{ELSEIF}  ...
\textbf{ELSEIF}   \s{C}{n}  \textbf{THEN} f = \s{res}{n}
\end{code}
\end{minipage}
\end{minipage}

\caption{Semantics of Horizontal Condition Table (HCT)}\label{fig:HCT}
\end{center}
\end{figure}
%
In documenting input-output behaviours using horizontal condition tables (HCTs), we need to reason about their \emph{completeness} and \emph{disjointness}. Suppose there is no sub-condition, completeness ensures that at least one row is applicable to every input, i.\,e.,~$(C_1 \vee C_2 \vee \dots \vee C_{n} \equiv \name{True})$. Disjointness ensures that rows do not overlap, e.\,g.,~($i \neq j \Rightarrow \lnot(C_i \land C_j)$). Similar constraints apply to the sub-conditions, if any.  

\smallskip

\noindent\textbf{Choice of Theorem Prover.} We chose the PVS theorem prover to formalize the input-output requirements of function blocks primarily because it supports the syntax and semantics of tables. In particular, for each table that is syntactically valid, PVS automatically generates its associated healthiness conditions of completeness and disjointness as type correctness conditions (TCCs). Furthermore, we have expertise built from past experience in applying PVS to check requirements and designs in the nuclear domain~\cite{LMFM00} that gave us confidence in using the toolset. For modelling real-time behaviour, we reused parts of the PVS theories from \cite{HuLawfordWassyng2008,Hu08} (see Sec.~\ref{preliminaries:software_samples} to \ref{preliminaries:timer_impelementation}). 

For presentation, we show PVS listings using ASCII characters in frame boxes, whereas in the main text, we typeset names of predicates, types, theorems, \emph{etc}., in the math form.

\smallskip

\npar{Modelling Time in the Physical Domain} 

As PLCs are widely used in real-time systems, the modelling of time is critical. For our purpose of verification, we approximate the continuous time in the physical domain as a type \name{tick}, defined as a discrete series of equally-distributed clock ticks, with an arbitrarily small positive time interval $\delta$ between two consecutive clock ticks: $\textit{tick} = \{ t_n : \mathbb{R}_{\ge 0} \mid \delta \in \mathbb{R}_{> 0} \land (\exists n : \mathbb{N} \bullet t_n = n \times \delta) \}$. We also define \name{not\_init}, a subtype of \name{tick} that excludes $t_0$. We define operators to manipulate values at the tick level: $init(t: tick) = (t = 0)$, $pre(t: not\_init) = (t - \delta)$, $next(t: tick) = (t + \delta)$, and $rank(t: tick) = \frac{t}{\delta}$. We often apply induction to prove properties that should hold over time\footnote{\texttt{pred[tick]} is synonymous to the function type \texttt{[tick -> bool]}}:
%
\begin{pvs}
time_induction: THEOREM
  FORALL (P: pred[tick]):   
     (FORALL (t: tick): init(t) => P(t)) 
   & (FORALL (t: not_init): P(pre(t)) => P(t)) => (FORALL (t: tick): P(t))
\end{pvs}

\noindent where $pred[tick]$ is a PVS shorthand for ``predicates on tick'' (i.e., functions mapping $tick$ to Boolean).

\smallskip

\npar{Modelling Samples in the Software Domain}\label{preliminaries:software_samples} 

\begin{figure}[htb]
\begin{center}
\includegraphics[width=.5\textwidth]{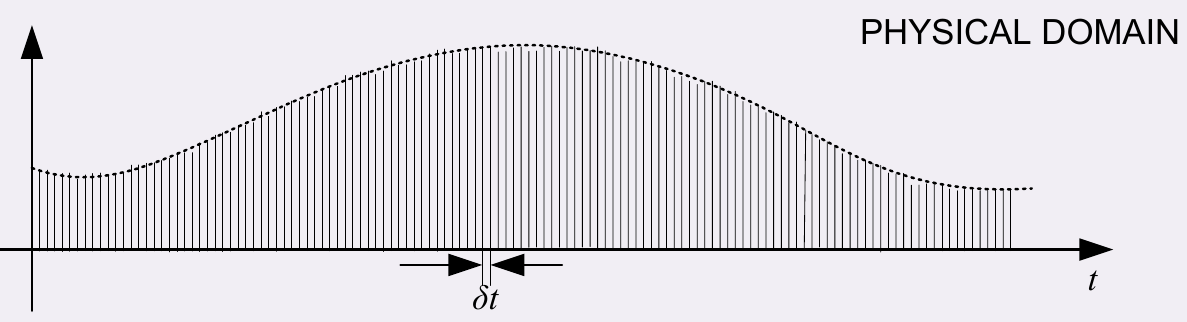}
\includegraphics[width=.49\textwidth]{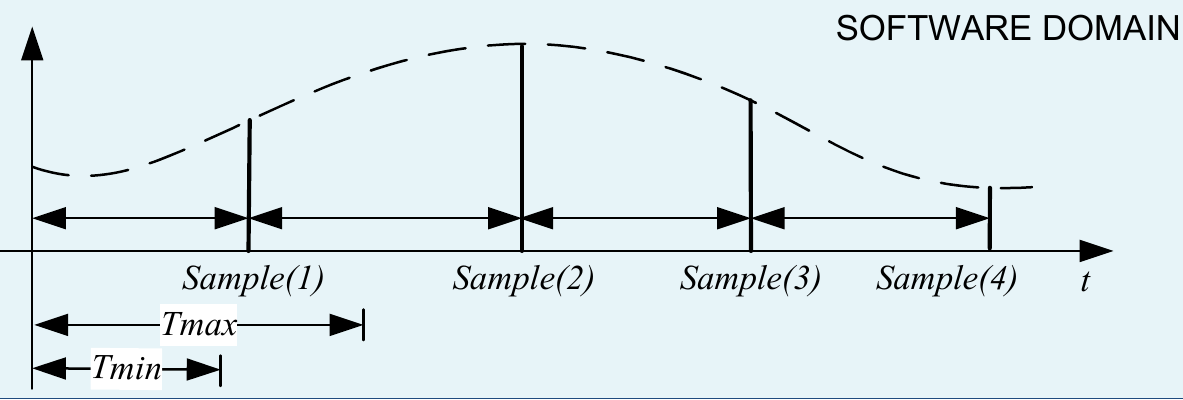}

\caption{Clock Ticks in Physical Domain vs. Samples in Software Domain~\cite[p81]{Hu08}}\label{fig:ClockTicks}
\end{center}
\end{figure}

We use a variable $\textit{Sample}: \mathbb{N} \rightarrow \mathbb{R}_{\geq 0}$ to denote the series of samples over time, such that the time of each sample (i.e., $\textit{Sample(n)}$, $n \in \mathbb{N}$) maps to a valid clock tick. As shown in Fig.~\ref{fig:ClockTicks}, realistically, the clock tick frequency $\frac{1}{\delta}$ in the physical domain should be significantly larger than the sampling frequency in the software domain. We bound sample intervals between \name{Tmin} and \name{Tmax}, determined by considering the shortest time after which events must be detected.

As rates of clock ticks and sampling are distinct, a monitored signal \name{Pf} that rapidly changes between two consecutive samples (called a ``spike'') can cause inconsistent results produced in the two domains. To rule out such scenarios, we define a predicate subtype $\mathit{FilteredTickPred}$\footnote{An example of using the subtype $\mathit{FilteredTickPred}$ to constrain input signals can be found in the verification story of the \name{Pushbutton} subsystem (Sec.~\ref{sec:PB}).} that only allows monitored conditions which remain unchanged between consecutive samples: 
%
\begin{pvs}
  FilteredTickPred?(P: pred[tick]): bool =
        ( FORALL t0: P(t0) /= P(next(t0)) =>
            (FORALL (t: tick): 
               t0 < t AND t <= t0 + Tmax => P(next(t0)) = P(t)) )
    AND ( FORALL (t: tick):
             t <= Tmax => P(t) = P(0) )

  FilteredTickPred: TYPE+ = (FilteredTickPred?)
  Pf: VAR FilteredTickPred
\end{pvs}

\smallskip

\npar{Operators for Specifying Timing Requirements} 

As discussed in Sec.~\ref{sec:intro}, we define the infix operator: 
\[
Held\_For: (tick \rightarrow \mathbb{B}) \times (tick \rightarrow \mathbb{R}_{> 0}) \times (tick \rightarrow \mathbb{R}_{\geq 0}) \times (tick \rightarrow \mathbb{R}_{\geq 0}) \rightarrow (tick \rightarrow bool)
\]
to specify a common functional timing requirement, e.g., $\textit{P}\ \mathbf{Held\_For}\ (\textit{d}, \delta \textit{L}, \delta \textit{R})$, that a monitored boolean condition \name{P} should sustain over a positive time duration \name{d}, with non-negative left tolerance $\delta \textit{L}$ and right tolerance $\delta \textit{R}$. More precisely,
\[
\textit{P}\ \mathbf{Held\_For}\ (\textit{d}, \delta \textit{L}, \delta \textit{R})(t_{now}) \equiv (~\exists t_{j}: t_{now} - t_{j} \ge \textit{d} \; \bullet \; (\forall t_{i}: t_{j} \le t_{i} \le t_{now} \bullet P(t_{i}))~)
\]
\noindent where $\textit{d} \in [\textit{d}(t_{now})-\delta \textit{L}(t_{now}), \textit{d}(t_{now})+\delta \textit{L}(t_{now})]$. In our model of time, inputs and outputs are represented as functions mapping ticks to values. For example, the left tolerance may change from $\delta \textit{L}(t_1)$ to $\delta \textit{L}(t_2)$. However, as discussed in Sec.~\ref{sec:intro}, the behaviour of \name{Held\_For} is nondeterministic when \name{P} has held $\mathit{TRUE}$ for a period that is bounded by $[d-\delta L, d+\delta R]$.

To resolve the non-determinism in \name{Held\_For}, we define two refinement operators: \name{Held\_For\_S} and \name{Held\_For\_I}. Both operators are deterministic by fixing the duration \name{d} in the above definition of \name{Held\_For} as $\textit{d}(t_{now})-\delta \textit{L}(t_{now})$. We will only see \name{Held\_For\_I} in the case studies, but it is defined in terms of \name{Held\_For\_S}. \name{Held\_For\_S} is a partial function on \name{tick} that produces values only at points of sampling (i.e., it is undefined on ticks in-between samples). 
%
\begin{pvs}
  Held_For_S(P, duration, Sample)(ne): bool =
    EXISTS (n0 : nat): 
          Sample(ne) - Sample(n0) >= duration
      AND FORALL (n: nat): n0 <= n AND n <= ne => P(Sample(n))
\end{pvs}
%
\noindent On the other hand, \name{Held\_For\_I} is a totalized version of \name{Held\_For\_S}: its value at time \name{t}, where $\textit{Sample(n)} \leq \textit{t} < \textit{Sample(n + 1)}$, is equivalent to that produced at time \name{Sample(n)} (i.e., the closest left sample calculated by $\mathit{Left\_Sample}$).
%
\begin{pvs}
  Held_For_I(P, duration, Sample)(t): bool = 
    Held_For_S(P, duration, Sample)(Left_Sample(Sample, t))
\end{pvs}
%

\smallskip

\npar{Implementing the Held\_For\_I Timing Operator}\label{preliminaries:timer_impelementation}

We use \name{Timer\_I} (defined in terms of \name{Timer\_S}) to implement the \name{Held\_For\_I} timing operator. \name{Timer\_I} agrees on outputs from \name{Timer\_S} at sample points and keep the same value at any clock tick until the next sample point (this is analogous to how \name{Held\_For\_I} is related to \name{Held\_For\_S}). 
%
\begin{pvs}
  Timer_I(P, Sample, TimeOut)(t): tick = 
    Timer_S(P, Sample, TimeOut)(Left_Sample(Sample, t))
\end{pvs}
%
\noindent where \name{Timer\_S}\cite[p. 97]{Hu08} counts, starting from the closest left sample to the clock tick in question, for how long the monitored condition \name{P} has been enabled, and stops counting when \name{TimeOut} is reached. The output type of \name{Timer\_S} is \name{tick}, calculated from how many samples \name{P} has been held across. As mentioned in Sec.~\ref{sec:intro}, the theorem \name{TimerGeneral\_I} is proved~\cite[p. 99]{Hu08} to ensure that \name{Timer\_I} is a proper implementation for \name{Held\_For\_I}.

\smallskip

\npar{A Formal Approach to Specifying and Verifying Function Blocks}\label{sec:approach}

Our reported approach~\cite{PangWangLawfordWassyng2013} fits into the timing model as described above. For each FB, its input-output requirements and FBD implementation are formalized in PVS as two (higher-order) predicates, parameterized by input and output lists. Each input or output is represented as a timed sequence (or trajectory) mapping clock ticks to values (e.g., $[tick \rightarrow real]$). Without loss of generality we write $i$ and $o$ to denote, respectively, the lists of input and output trajectories.

Consider a composite function block $\mathit{FB}$ (e.g., see Fig.~\ref{fig:Sealedin_impl} in Sec.~\ref{sec:Sealin}). The \emph{requirements predicate} of $\mathit{FB}$ (denoted as \name{FB\_REQ}, e.g., \name{Trip\_sealedin\_REQ}) returns true if its outputs are related to inputs in the expected way (specified using tabular expressions) across all time ticks. The \emph{implementation predicate} of $\mathit{FB}$ (denoted as \name{FB\_IMPL}, e.g., \name{Trip\_sealedin\_IMPL}) is constructed by composing, using logical conjunction, the requirements predicates of its component FBs (e.g., \name{TON}, \name{CONJU}, \emph{etc}.) as configured in its FBD implementation. All inter-connectives (e.g., \name{w1}, \name{w2}, \emph{etc}.) in the FBD implementation are hidden using an existential quantification. 

\smallskip

\noindent\textbf{Proof of Consistency} To ensure that the implementation is consistent or feasible, we prove that for each list of input trajectories, there exists at least one list of output trajectories such that \name{FB\_IMPL} is defined:
\begin{equation}\label{consistency}
\begin{split}
\vdash\forall i \; \bullet \; \exists o\; \bullet \; 
\textit{FB\_IMPL}(i,\; o)
\end{split}
\end{equation}

\smallskip

\noindent\textbf{Proof of Correctness} To ensure that the implementation is correct with respect to the intended requirement, we prove that the observable inputs and outputs conform to those of the requirements:
\begin{equation}\label{correctness}
\begin{array}{l}
\vdash\forall i \; \bullet \; \forall o \; \bullet 
 FB\_IMPL(i,\; o)
\Rightarrow FB\_REQ(i,\; o)
\end{array}
\end{equation}

%% file: input_sec_ton_with_tolerances.tex
\section{Formalizing IEC~61131-3 Timers with Tolerances}\label{sec:timers}

We present the first contribution of this paper: incorporating the notion of \emph{timing tolerances}~\cite{Wassyng05} (i.e., the controller's reaction to the environment is associated with a delay) into the formalization of the black-box, input-output requirements of IEC~61131-3 timers. Such formalization improves the accuracy of our previous work~\cite{PangWangLawfordWassyng2013} by making the resulting requirements models \emph{implementable}. 

In IEC~61131-3 there are three timer FBs: \name{TON} (On-delay), \name{TOF} (Off-delay), and \name{TP} (Pulse) timers. As case studies presented in this paper (Sec.~\ref{sec:Sealin} and Sec.\ref{sec:PB}) only make use of the \name{TON} block, in this section we present its re-formalization only and report details of the other two timer blocks in~\cite{PangWangLawfordWassyngTechReport2014}.

\input{figs/input_fig_ton_decl_timing_diagram}

The \name{TON} block is commonly used as a component of safety-critical systems. For example, it can be used to determine if a sensor signal has gone out of its safety range for too long, as we will see in Sec.~\ref{sec:Sealin} and Sec.~\ref{sec:PB}. Fig.~\ref{fig:ton} shows, extracted from IEC~61131-3, the input-output declaration (on the LHS) and a timing diagram\footnote{The horizontal axis is labelled with time instants $t_{i}$, $i \in 0..5$} (on the RHS) illustrating the expected behaviour of the \name{TON} block. The \name{TON} block is declared with two inputs (a boolean condition \name{IN} and a time period of length \name{PT}) and two outputs (a boolean value \name{Q} and a length \name{ET} of time period). Timer \name{TON} monitors the input condition \name{IN} and sets the output \name{Q} as true whenever \name{IN} remains enabled for longer than a time period of some input length \name{PT}. If the monitored input \name{IN} has been enabled for some time $t < PT$, then the timer sets the output \name{ET} (i.e., elapsed time) with value \name{t}; otherwise, it sets \name{ET} with value \name{PT}. 

The use of a timing diagram by IEC~61131-3 to describe the expected behaviour of the \name{TON} block (and the other two timers) is limited to an incomplete set of use cases. As a result, we attempted in~\cite{PangWangLawfordWassyng2013} to use function tables to formalize the black-box, input-output requirements of the three timer blocks (on-delay, off-delay, and pulse timers) listed in IEC~61131-3. Fig.~\ref{fig:ton_req} shows our previous attempt of the requirements specification of the \name{TON} block, where $t$ denotes the current clock tick, and a time stamp \name{last\_enabled} is used to record the exact time (with no delay) that the input condition \name{IN} just becomes enabled. However, the requirements model in Fig.~\ref{fig:ton_req} is not implementable because it describes idealized behaviour: the timer (or the controller) reacts instantaneously to changes in the environment.

\input{figs/input_fig_ton_tab_idealized.tex}


As part of the contribution of this paper, we revise the function tables of all three timers in IEC~61131-3 by incorporating the notion of timing tolerances~\cite{Wassyng05}. To achieve this, we use the pre-verified operator $Timer\_I$ (Sec.~\ref{sec:preliminary}) to redefine requirements of the three timers (e.g., Fig.~\ref{fig:ton_impl} for the $\mathit{TON}$ timer).

\input{figs/input_fig_ton_tab_tolerances.tex}

The essence of our first contribution presented in this section is that we incorporate the notion of timing tolerances, via the use of the pre-verified operator $\mathit{Timer\_I}$, into the requirements of IEC~61131 timers so that they are implementable. This allows us to conduct case studies such as the one in Sec.~\ref{sec:Sealin} on implementing and verifying subsystems using the IEC~61131-3 timers.

%% file: figs/input_fig_ton_decl_timing_diagram.tex
\begin{figure}[h]
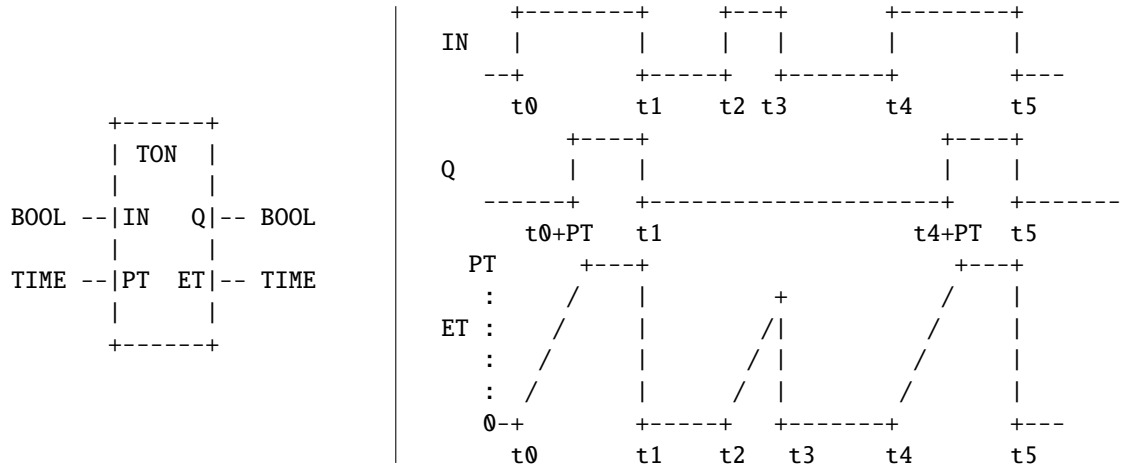

\begin{minipage}{\textwidth}

\begin{tabular}{c|c}
\begin{minipage}{.3\textwidth}
\begin{code}
       +------+
       | TON  |
       |      |
BOOL --|IN   Q|-- BOOL
       |      |
TIME --|PT  ET|-- TIME
       |      |
       +------+
\end{code}
\end{minipage}
\quad
&
\quad
\begin{minipage}{.7\textwidth}
\begin{code}
     +--------+     +---+       +--------+
IN   |        |     |   |       |        |
   --+        +-----+   +-------+        +---
     t0       t1    t2 t3       t4       t5
         +----+                     +----+
Q        |    |                     |    |
   ------+    +---------------------+    +-------
      t0+PT   t1                  t4+PT  t5   	
  PT      +---+                      +---+
   :     /    |         +           /    |
ET :    /     |        /|          /     |
   :   /      |       / |         /      |
   :  /       |      /  |        /       |
   0-+        +-----+   +-------+        +---
     t0       t1    t2   t3     t4       t5
\end{code}
\end{minipage}
\end{tabular}

\end{minipage}
\caption{\name{TON} timer declaration and definition in timing diagram~\cite{IEC:2003:IEP}}
\label{fig:ton}
\end{figure}

%% file: figs/input_fig_ton_tab_idealized.tex
\begin{figure}[h]\centering
\begin{tabular}{|cc}
\multicolumn{1}{c}{}
&  \multicolumn{1}{c}{\small{\textit{Result}}}\\\hhline{~-}
\multicolumn{1}{c}{\small{\textit{Condition}}}&\multicolumn{1}{||c||}{\cellcolor[gray]{0.8}last\_enabled}\\ \cline{1-2}
$\neg$IN$_{-1}$ $\wedge$ IN & \multicolumn{1}{||c||}{t} \\ \cline{1-2}
IN$_{-1}$ $\vee$ $\neg$IN & \multicolumn{1}{||c||}{NC} \\ \cline{1-2}
\end{tabular}\quad
\begin{tabular}{|cc}
\multicolumn{1}{c}{}
&  \multicolumn{1}{c}{\small{\textit{Result}}}\\\hhline{~-}
\multicolumn{1}{c}{\small{\textit{Condition}}}&\multicolumn{1}{||c||}{\cellcolor[gray]{0.8}Q}\\ \cline{1-2}
IN $\wedge$ (d $\ge$ PT) & \multicolumn{1}{||c||}{TRUE} \\ \cline{1-2}
IN $\wedge$ (d $\lt$ PT) & \multicolumn{1}{||c||}{FALSE} \\ \cline{1-2}
$\neg$IN & \multicolumn{1}{||c||}{FALSE} \\ \cline{1-2}
\end{tabular}\quad
\begin{tabular}{|cc}
\multicolumn{1}{c}{}
&  \multicolumn{1}{c}{\small{\textit{Result}}}\\\hhline{~-}
\multicolumn{1}{c}{\small{\textit{Condition}}}&\multicolumn{1}{||c||}{\cellcolor[gray]{0.8}ET}\\ \cline{1-2}
IN $\wedge$ (d $\ge$ PT) & \multicolumn{1}{||c||}{PT} \\ \cline{1-2}
IN $\wedge$ (d $\lt$ PT) & \multicolumn{1}{||c||}{d} \\ \cline{1-2}
$\neg$IN & \multicolumn{1}{||c||}{0} \\ \cline{1-2}
\end{tabular}

\smallskip
\hfill \textbf{where} d stands for duration, d = t - last\_enabled 
\caption{Tabular Requirements of Timer \name{TON}: Idealized Behaviour}
\label{fig:ton_req}
\end{figure}

%% file: figs/input_fig_ton_tab_tolerances.tex
\begin{figure}[h]\centering
\begin{tabular}{|cc}
\multicolumn{1}{c}{}
&  \multicolumn{1}{c}{\small{\textit{Result}}}\\\hhline{~-}
\multicolumn{1}{c}{\small{\textit{Condition}}}&\multicolumn{1}{||c||}{\cellcolor[gray]{0.8}Q}\\ \cline{1-2}
d $\ge$ PT & \multicolumn{1}{||c||}{TRUE} \\ \cline{1-2}
d $\lt$ PT & \multicolumn{1}{||c||}{FALSE} \\ \cline{1-2}
\end{tabular}
\qquad
\begin{tabular}{|cc}
\multicolumn{1}{c}{}
&  \multicolumn{1}{c}{\small{\textit{Result}}}\\\hhline{~-}
\multicolumn{1}{c}{\small{\textit{Condition}}}&\multicolumn{1}{||c||}{\cellcolor[gray]{0.8}ET}\\ \cline{1-2}
d $\ge$ PT & \multicolumn{1}{||c||}{PT} \\ \cline{1-2}
d $\lt$ PT & \multicolumn{1}{||c||}{d} \\ \cline{1-2}
$\neg$IN & \multicolumn{1}{||c||}{0} \\ \cline{1-2}
\end{tabular}

\smallskip
\hfill \textbf{where} d stands for duration, d = (IN)~Timer\_I~(PT, $\delta$L, $\delta$R)
\caption{Tabular Requirements of Timer \name{TON}: Timing Tolerances Incorporated}
\label{fig:ton_impl}
\end{figure}

%% file: input_sec_case_study_trip_sealin.tex
\section{Case Study 1: the \name{Trip Sealed-In} Subsystem}\label{sec:Sealin}

In this section we apply our approach (Sec.~\ref{sec:approach}) to verify a candidate FBD implementation for the \name{Trip Sealed-In} subsystem. We identify an initialization error and suggest a fix.

\smallskip

\npar{Input-Output Declaration and Informal Description} 

The figure below declares the inputs and outputs of the \name{Trip Sealed-In} subsystem:
\begin{code}
                      +----------------------------------+
                      |          Trip Sealed-In          |
                      |                                  |
               BOOL --|Any_parm_trip                     |                     
\{e_Trip, e_NotTrip\} --|Trip                 Trip_SealedIn|-- BOOL
               REAL --|k_Sealindelay                     |
               BOOL --|Man_reset_req                     |
                      +----------------------------------+
\end{code}
\noindent \name{Trip Sealed-In} is a generic subsystem which monitors: 1) a set of sensor values; and 2) an alarm value produced by some other subsystem. It signals an alarm (denoted by the output \name{Trip\_SealedIn}), which may be manipulated by other subsystems, when two conditions are met. First, any of the monitored sensor values goes out of its safety range (called a parameter trip and denoted by an input condition \name{Any\_parm\_trip}). Second, the monitored input alarm is signalled continuously for longer than some preset constant \name{k\_Sealindelay}\footnote{The \name{k\_} name prefix is reserved for system-wide constants.} amount of time (denoted by an input value \name{Trip} of enumerated type $\{\name{e\_Trip}, \name{e\_NotTrip}\}$). Once the alarm \name{Trip\_SealedIn} is activated, it is not deactivated until all monitored sensor values fall back within their safety ranges, and then a manual reset is requested (denoted as an input \name{Man\_reset\_req}).

\smallskip

\npar{Tabular Requirements Specification with Timing Tolerances} We use a function table (Fig.~\ref{fig:Sealedin_req}) to perform a complete and disjoint analysis on the input domains. To incorporate timing tolerances into the requirements of \name{Trip Sealed-In}, we use the non-deterministic \name{Held\_For} operator (Sec.~\ref{sec:preliminary}) to specify a sustained window of duration $[\textit{k\_Sealindelay} - \delta \textit{L}, \textit{k\_Sealindelay} + \delta \textit{R}]$. 

\begin{figure}[!h]
\centering
\begin{tabular}{|c|cc}
\multicolumn{2}{c}{}
&  \multicolumn{1}{c}{\small{\name{Result}}}\\\hhline{~~-}
\multicolumn{2}{c}{\small{\name{Condition}}}&\multicolumn{1}{||c||}{\cellcolor[gray]{0.8}\name{Trip\_SealedIn}}\\ 
\cline{1-3}
\multirow{2}{*}{\name{Any\_parm\_trip}}&
(\name{Trip}=\name{e\_Trip})~\textbf{Held\_For}~(\name{k\_Sealindelay},\, $\delta$\name{L},\, $\delta$\name{R}) & \multicolumn{1}{||c||}{TRUE}\\ \cline{2-3}
& $\lnot$[(\name{Trip}=\name{e\_Trip})~\textbf{Held\_For}~(\name{k\_Sealindelay},\, $\delta$\name{L},\, $\delta$\name{R})] & \multicolumn{1}{||c||}{NC} \\ \hline
\multirow{2}{*}{$\neg$\name{Any\_parm\_trip}}&
\name{Man\_reset\_req} & \multicolumn{1}{||c||}{FALSE}\\ \cline{2-3}
&$\neg$\name{Man\_reset\_req} & \multicolumn{1}{||c||}{NC} \\ \hline
\end{tabular}
\caption{\name{Trip Sealed-In}: (non-deterministic) Requirements of  with Tolerances}\label{fig:Sealedin_req}
\end{figure}

However, for the purpose of verification in PVS, we reformulate the non-deterministic behaviour of Fig.~\ref{fig:Sealedin_req} in a recursive function\footnote{For proving termination, its progress is measured using discrete time instants \name{rank(t)}.} using the deterministic \name{Held\_For\_I} operator to impose the constraint that only a single value (i.e., $\textit{k\_Sealindelay} - \textit{delta\_L}$ where both are declared constants) is chosen from the duration and is used consistently for detecting sustained events.

Below we define a recursive function \name{Trip\_SealedIn\_f} over all clock ticks:

\begin{pvs}
Trip_SealedIn_f(Any_parm_trip: pred[tick], 
                Trip         : [tick->{e_Trip, e_NotTrip}],
                Man_reset_req: pred[tick])(t: tick)
: RECURSIVE bool =	
  IF init(t) THEN TRUE ELSE 
  LET
   TRIPPED = LAMBDA (t: tick): Trip(t) = e_Trip,
   HELD    = Held_For_I(TRIPPED,k_Sealindelay-delta_L,Sample)(t),
   PREV    = Trip_SealedIn_f(
                Any_parm_trip,Trip,Man_reset_req)(pre(t))        
  IN TABLE
     $\mid$     Any_parm_trip(t) &     HELD             $\mid$ TRUE     $\mid\mid$
     $\mid$     Any_parm_trip(t) & NOT HELD             $\mid$ PREV     $\mid\mid$
     $\mid$ NOT Any_parm_trip(t) &     Man_reset_req(t) $\mid$ FALSE    $\mid\mid$  
     $\mid$ NOT Any_parm_trip(t) & NOT Man_reset_req(t) $\mid$ PREV     $\mid\mid$   
  ENDTABLE ENDIF MEASURE rank(t)
\end{pvs}

Using \name{Trip\_SealedIn\_f}, we have deterministic requirements (Fig.~\ref{fig:Sealedin_req_f}) for the \name{Trip Sealed-In} subsystem:

\begin{figure}[h]
\begin{pvs}
Trip_SealedIn_REQ(Any_parm_trip: pred[tick], 
                  Trip         : [tick->{e_Trip, e_NotTrip}], 
                  Man_reset_req: pred[tick], 
                  TripSealedIn : pred[tick]): bool 
= FORALL (t: tick):
    TripSealedIn(t) = 
      Trip_SealedIn_f(Any_parm_trip, Trip, Man_reset_req)(t)
\end{pvs}
\caption{\name{Trip Sealed-In}: (deterministic) Requirements of  with Tolerances in PVS}\label{fig:Sealedin_req_f}
\end{figure}
\textit{Remark}. Compared with Fig.~\ref{fig:Sealedin_req}, the use of the operator \name{Held\_For\_I} in Fig.~\ref{fig:Sealedin_req_f} resolves the non-determinism by fixing the level of tolerance (i.e., as the alarm input \name{Trip} has been activated for or longer than $\textit{k\_Sealindelay}-\delta \textit{L}$, the \name{Trip Sealed-In} subsystem is guaranteed to detect it and act accordingly).

\smallskip

\begin{figure}[h]
\centering
\includegraphics[width=\textwidth]{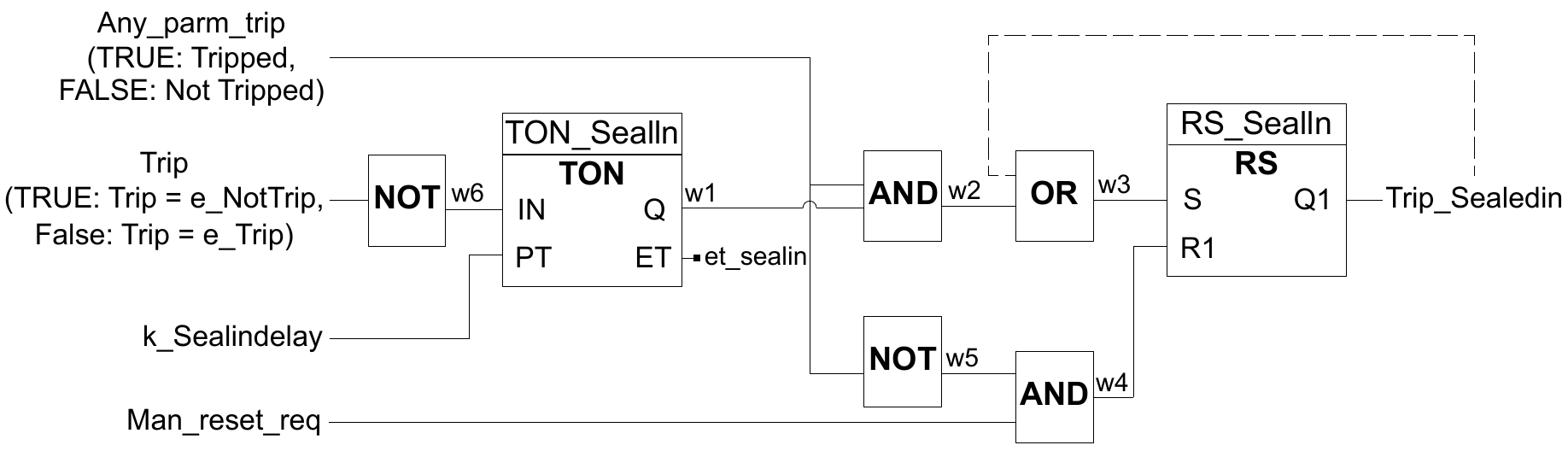}
\caption{\name{Trip Sealed-In} implementation in FBD}
\label{fig:Sealedin_impl}
\end{figure}

\npar{Formalizing the FBD Implementation} We propose a FBD implementation (Fig.~\ref{fig:Sealedin_impl}) which should satisfy the requirements (Fig.~\ref{fig:Sealedin_req_f}). 

We use the IEC~61131 \name{TON} timer (see Sec.~\ref{sec:timers} for its formalization incorporated with tolerances) to implement the use of the \name{Held\_For\_I} operator (subject to a correctness proof which we will discuss below). As the recursive function used to define the requirements depends on the value of itself (at the previous time tick), we specify a feedback loop (dashed line) in the implementation. 

The use of the left-most \name{NOT} (negation) block in Fig.~\ref{fig:Sealedin_impl} has to do with the mismatch between types at the requirements level (i.e., $\{\name{e\_Trip}, \name{e\_NotTrip}\}$) and that at the FB implementation level (i.e., boolean): somehow the engineers interpret value \name{e\_Trip} as \name{FALSE} and \name{e\_NotTrip} as \name{TRUE}, so a conversion is necessary to make sure the \name{Trip Sealed-In} has a consistent interpretation. The requirements that the alarm output \name{Trip\_Sealedin} is deactivated (or reset) when there is no parameter trips, and when a manual reset is requested, is implemented using a standard block \name{RS} (reset dominant flip flop).

To prove that the proposed FBD implementation of \name{Trip Sealed-In} (Fig.~\ref{fig:Sealedin_impl}) is both feasible and conforms to its requirements (Fig.~\ref{fig:Sealedin_req_f}), we follow our approach (Sec.~\ref{sec:approach}) to formalize it by composing, using conjunction, the formalizing predicates\footnote{Predicates \name{NOT} (logical negation), \name{CONJ} (logical conjunction), \name{DISJ} (logical disjunction), \name{TON} (on-delay timer), and \name{RS} (reset dominant latch).} of all component blocks (all inter-connectors are hidden using an existential quantification.):
\[
\begin{array}{l}
\textit{Trip\_sealedin\_IMPL}(Any\_parm\_trip, Trip,  Man\_reset\_req, Trip\_SealedIn) \\
\quad \equiv \exists \; w_{1},w_{2},w_{3}, w_{4},w_{5},w_{6}, et\_sealin \; \bullet \\ 
\qquad\quad \bigbrace{
\; \; \; \textit{NOT}(\textit{Trip}, w_{6})\\
\land \; \textit{TON}(w_{6}, k\_Sealindelay - \delta L, w_{1}, et\_sealin) \\
\land \; \textit{CONJ}(Any\_parm\_trip, w_{1}, w_{2})\\
\land \; \textit{DISJ}(w_{2},\textit{Trip\_SealedIn}, w_{3})\\

\land \; \textit{NOT}(\textit{Any\_parm\_trip}, w_{5}) \\
\land \; \textit{CONJ}(w_{5}, \textit{Man\_reset\_req}, w_{4})\\

\land \; \textit{RS}(w_{4}, w_{3}, \textit{Trip\_SealedIn})}
\end{array}
\]

\smallskip

\npar{Proofs of Consistency and Correctness} First, we prove that the FBD implementation (Fig.~\ref{fig:Sealedin_impl}) is feasible by instantiating formula~(\ref{consistency}) in Sec.~\ref{sec:approach}: 
\[
\begin{array}{l}
\vdash\forall \; \textit{Any\_parm\_trip}, \textit{Trip}, \textit{Man\_reset\_req} \; \bullet \\ 
\qquad \exists \; \textit{Trip\_SealedIn} \; \bullet \textit{Trip\_sealedin\_IMPL}(\\ \qquad
\qquad\begin{array}{l}
\textit{Any\_parm\_trip},\ \textit{AbstParmTrip\_timed(Trip)},\ 
\textit{Man\_reset\_req},\ \textit{Trip\_SealedIn})
\end{array}
\end{array}
\]
\noindent The abstraction function \name{AbstParmTrip\_timed} handles the mismatched types of input \name{Trip} at levels of requirements and implementation (e.g., \name{e\_NotTrip} mapped to \name{TRUE}). We discharge the consistency proof using proper instantiations.

Second, we prove that the FBD implementation is correct with respect to Fig.~\ref{fig:Sealedin_req_f}, considering timing tolerances, by instantiating formula~(\ref{correctness}) in Sec.~\ref{sec:approach}:
\[
\begin{array}{l}
\vdash\forall \; \textit{Any\_parm\_trip},\ \textit{Trip},\ \textit{Man\_reset\_req},\ \textit{Trip\_SealedIn} \; \bullet \\ 
\qquad \textit{Trip\_sealedin\_IMPL}(\textit{Any\_parm\_trip},\, \textit{AbstParmTrip\_timed(Trip)},\, \textit{Man\_reset\_req},\, \textit{Trip\_SealedIn}) \\
\qquad  \quad \Rightarrow \textit{Trip\_sealedin\_REQ}(\textit{Any\_parm\_trip}, \textit{Trip}, \, \textit{Man\_reset\_req},\, \textit{Trip\_SealedIn}) \\
\end{array}
\]
As there is a feedback loop in the FBD implementation (Fig.~\ref{fig:Sealedin_impl}), our strategy of discharging the correctness theorem is by mathematical induction (using the \name{time\_induction} proposition in Sec.~\ref{sec:preliminary}) over tick values. Since the \name{Timer\_I} operator (Sec.~\ref{sec:preliminary}) is used to formalize the requirements of the \name{TON} timer that contributes to the FBD implementation, its definition is expanded in both the base and inductive cases.

\begin{figure}[h]
\centering
\includegraphics[width=\textwidth]{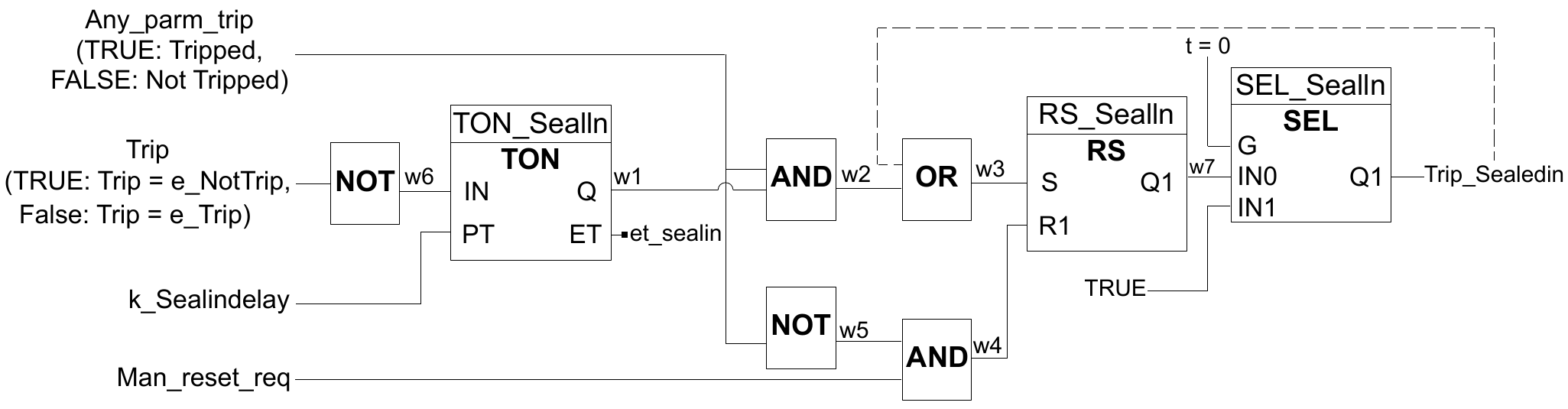}
\caption{Revised \name{Trip Sealed-In} implementation in FBD}
\label{fig:Sealedin_impl_revised}
\end{figure}

However, when proving the base case (when $\textit{t} = 0$), we found that the initial value of output \name{Q1} of the \name{RS\_Sealin} block and the initial value of the subsystem output \name{Trip\_SealedIn}~---~these two values are directly connected in the initial FBD implementation (Fig.~\ref{fig:Sealedin_impl})~---~are inconsistent. According to the SRS (Software Requirements Specification), the value of \name{Trip\_SealedIn} is initialized to \name{TRUE}, whereas that of \name{Q1} is \name{FALSE}. We resolve this issue of inconsistency by suggesting a revised FBD implementation (Fig.~\ref{fig:Sealedin_impl_revised}) and prove that it is correct with respect to Fig.~\ref{fig:Sealedin_req_f}. In this revised implementation, we add an IEC~61131-3 selection block \name{SEL\_Sealin}, acting as a multiplexer to discriminate the value of \name{Q1} (at the initial tick and at the non-initial tick) that is output as \name{Trip\_SealedIn}.

\textit{Remark.} We just illustrated that, by adopting our approach, we are able to justify the appropriateness of a candidate FBD implementation, and to fix it accordingly if necessary.

%

%% file: input_sec_case_study_pushbutton.tex
\section{Case Study 2: the \name{Pushbutton} Subsystem}\label{sec:PB}

In this section we apply our approach (Sec.~\ref{sec:approach}) to verify a candidate FBD implementation for the \name{Pushbutton} subsystem. We identify a missing assumption of implementation and suggest a solution.

\smallskip

\npar{Input-Output Declaration and Informal Description} The figure below declares the inputs and outputs of the \name{Pushbutton} subsystem. 
\begin{code}
                 +------------------------------------+
                 |            Pushbutton              |
                 |                                    |                
          y_pb --|m                                   |
          REAL --|k_Debounce              f_Pushbutton|-- y_pbdesign
          REAL --|k_Stuck                             |
                 +------------------------------------+
\end{code}
\noindent\name{Pushbutton} is a generic subsystem which monitors the status of a pushbutton (denoted by an input $\textit{m} \in \{\textit{e\_Pressed}, \textit{e\_NotPressed}\}$), which may be pressed to manually, e.g., enable or disable a sensor trip\footnote{A sensor trip occurs if the sensor signal in question goes above its set point.}. Its behaviour is denoted by an output $\textit{f\_Pushbutton} \in \{\textit{e\_pbNotDebounced}, \textit{e\_pbDebounced}, \textit{e\_pbStuck}\}$. \name{Pushbutton} determines if either: (a) the button is not pressed, or pressed but not for a sufficient period of time (denoted by some pre-set value \name{k\_Debounce}\footnote{The \name{k\_} name prefix is reserved for system-wide constants.}) to register as a press; (b) the button is pressed long enough to quality as a press; or (c) the button is pressed for longer than some pre-set period of time (denoted by \name{k\_Stuck}) without bouncing back and thus is considered stuck. 

\smallskip

\npar{Tabular Requirements Specification with Timing Tolerances} 

For the purpose of verification in PVS, we use the function table below\footnote{The PVS encoding of this table is not shown in this paper.} to perform a complete and disjoint analysis on the domain of the button status. To incorporate timing tolerances, similar to the requirements specification for the \name{Trip Sealed-In} subsystem (Fig.~\ref{fig:Sealedin_req_f}, p.\pageref{fig:Sealedin_req_f}), we use the deterministic \name{Held\_For\_I} operator (Sec.~\ref{sec:preliminary}), where values $\textit{k\_Debounce} - \delta \textit{L}$ and $\textit{k\_Stuck} - \delta \textit{L}$ are chosen and used consistently for detecting the sustained events.

\begin{center}
\begin{tabular}{|cc}
\multicolumn{1}{c}{}
&  \multicolumn{1}{c}{\small{\textit{Result}}}\\\hhline{~-}
\multicolumn{1}{c}{\small{\textit{Condition}}}&\multicolumn{1}{||c||}{\cellcolor[gray]{0.8}\name{f\_Pushbutton}}\\ \cline{1-2}
\name{m} = \name{e\_NotPressed} & \multicolumn{1}{||c||}{\name{e\_pbNotDebounced}} \\ \cline{1-2}
(\name{m} = \name{e\_Pressed}) $\land$ $\lnot$\name{debounced} & \multicolumn{1}{||c||}{\name{e\_pbNotDebounced}} \\ \cline{1-2}
\name{debounced} $\land$ $\lnot$\name{stuck}   & \multicolumn{1}{||c||}{\name{e\_pbDebounced}} \\ \cline{1-2}
\name{stuck}   & \multicolumn{1}{||c||}{\name{e\_pbStuck}} \\ \cline{1-2}
\multicolumn{2}{r}{
$\begin{array}{lll}
\textbf{where} & \textit{debounced} & = (\textit{m} = \textit{e\_Pressed})~\textbf{Held\_For\_I}~(\textit{k\_Debounce} - \delta \textit{L}) \\
               & \textit{stuck} & = (\textit{m} = \textit{e\_Pressed})~\textbf{Held\_For\_I}~(\textit{k\_Stuck} - \delta \textit{L})
\end{array}$} \\
\end{tabular}
\end{center}

\smallskip

\npar{Formalizing the FBD Implementation} We propose a FBD implementation which should satisfy the requirements:

\begin{center}
\centering
\includegraphics[width=.8\textwidth]{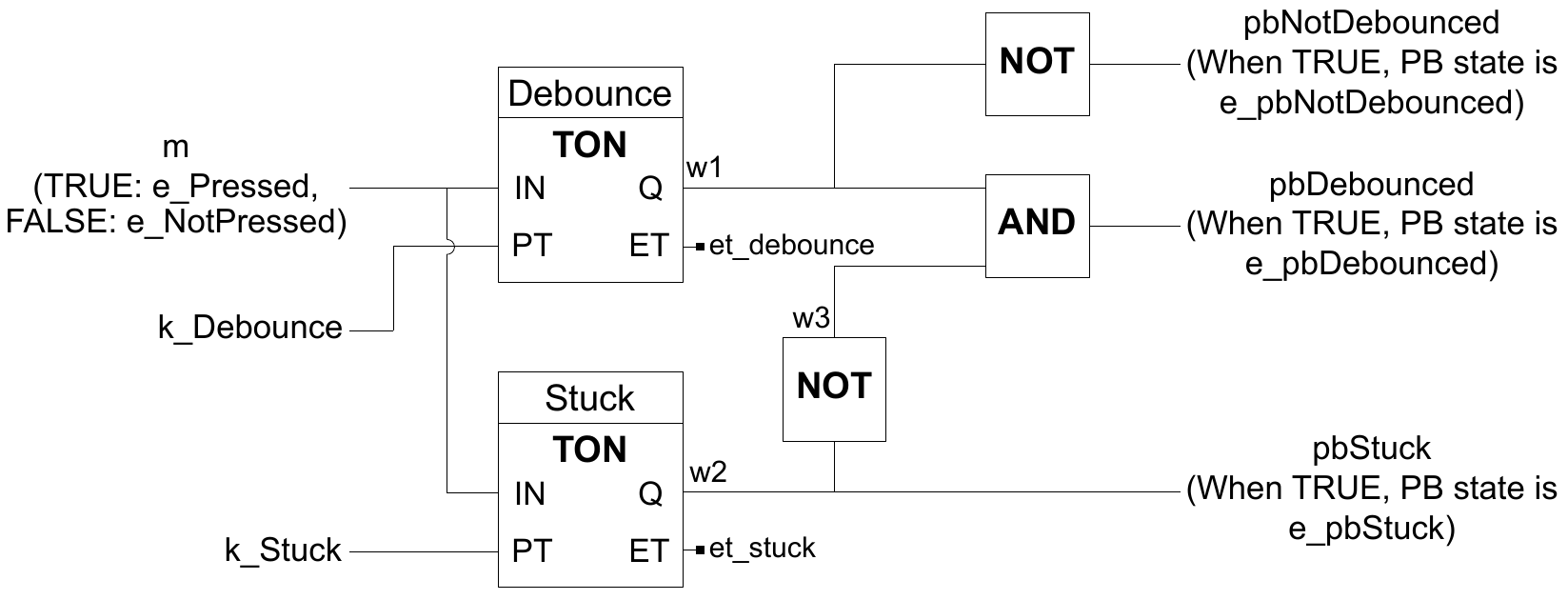}
\end{center}

\noindent We use two IEC~61131 \name{TON} timers (see Sec.~\ref{sec:timers} for its formalization) to implement the predicates \name{debounced} and \name{stuck} in the above requirements table that involve the use of the \name{Held\_For\_I} operator. Since only the button status is monitored, there is no need to specify a feedback loop in the implementation. To prove that this FBD implementation is consistent and correct, similar to what we do for that for the \name{Trip Sealed-In} subsystem (see Fig.~\ref{fig:Sealedin_impl}, p.\pageref{fig:Sealedin_impl}), we formalize it by composing the formalizing predicates of all its component blocks using conjunction, and by hiding inter-connectors using an existential quantification.

\smallskip

\npar{Proof Obligations: Consistency and Correctness} The consistency and correctness theorems for the \name{Pushbutton} subsystem are stated in a similar manner as those for the \name{Trip Sealed-In} subsystem by properly instantiating, respectively, formulas~\ref{consistency} and \ref{correctness} in Sec.~\ref{sec:approach}. However, we had difficulties when first attempting to prove that the above requirements table for \name{f\_Pushbutton} possesses the disjointness property. To resolve this, we tried to simplify the requirements table by collapsing the first two rows into a single one with the input condition $\lnot pressed \land \lnot stuck$. This is done based on the observations that both row 1 and row 2 map to the same output value \name{e\_pbNotDebounced}, and that $\textit{m} = \textit{e\_Pressed} \lor \textit{m} = \textit{e\_NotPressed} \equiv \textit{true}$.

When proving that the revised requirements table is equivalent to the original one, we found a problematic scenario where the value of output \name{f\_Pushbutton} is produced inconsistently at the requirements and implementation levels: when the input condition \name{m} varies rapidly and generates a ``spike'', whose duration is shorter than the timing resolution. To rule out the ``spike'' scenarios for input \name{m}, we added an assumption, at the FBD implementation level, using the predicate subtype \name{FilteredTickPred} (Sec.~\ref{sec:preliminary}).

Finally, the revised requirements table can be proved for its completeness, disjointness, consistency, and correctness by following a similar pattern of proofs as for the \name{Trip Sealed-In} subsystem. For proving the correctness theorem, as there is not a feedback loop in the above FBD implementation, we do not need to discharge the correctness theorem using mathematical induction. Furthermore, as the \name{TON} components in the FBD implementation are formalized using the \name{Timer\_I} operator (Sec.~\ref{sec:timers}), we need to reuse the theorem \name{TimerGeneral\_I} with proper instantiations to show their equivalence to the \name{Held\_For\_I} expressions in the revised requirements table. 

%% file: input_sec_lessons.tex
\section{Proof Structure}\label{sec:lessons}

In the industrial software control system that we consider for this paper, the \name{Trip Sealed-In} subsystem implemented using a feedback loop (Sec.~\ref{sec:Sealin}) and the \name{Pushbutton} subsystem (Sec.~\ref{sec:PB}) are representative\footnote{This judgement is based on the use of a generic timing function, the \textit{Held\_For} operator, in the tabular expressions that describe the required behaviour.} of functionality in which FBD implementations make use of IEC~61131 timer blocks. Structures of their consistency and correctness proofs shall guide the proofs for many other subsystems of a similar nature. 
%

%

For illustration, we consider the correctness proof structure for the \name{Trip Sealed-In} subsystem. In principle, there are eight key steps to discharge the correctness theorem for a real-time subsystem implemented with a feedback loop (e.g., \name{Trip Sealed-In}). Except for the fourth step, where the \name{time\_induction} theorem is used to handle the feedback loop, others are standard commands.

1) Apply \verb|skosimp| to eliminate the universal quantification over input and output variables, and then apply \verb|flatten| to simplify theorem structure $\textit{impl} \Rightarrow \textit{req}$ by moving \name{impl} to the antecedent and \name{req} to the consequent. 2) Apply multiple \verb|expand| commands to unfold definitions of the requirements and implementation predicates. 3) In the antecedent, apply \verb|skolem!| to eliminate the existential quantification over inter-connectors. 4) To handle the recursive feedback loop, use the theorem \name{time\_induction} on $\textit{t} \in \textit{tick}$. 5) Apply a series of basic commands to complete the proof for the base case. 6) To prove the inductive case, first apply \verb|skolem!| and then \verb|expand| to unfold the recursive function that is used to define the requirements predicate (e.g., see Fig.~\ref{fig:Sealedin_req_f}, p.\pageref{fig:Sealedin_req_f}). 7) Apply \verb|split| and \verb|lift-if| to generate sub-goals. 8) Repeatedly apply: \verb|expand| commands to unfold definitions of the predicates for internal components, theorem \name{TimerGeneral\_I} with proper instantiations to link between \name{Held\_For\_I} in the requirements and \name{Timer\_I} in the implementation, and basic commands to complete the proof for the inductive step.

%% file: input_sec_related_works.tex
\section{Related Work}\label{sec:related-works}

The focus of this paper is the practical verification of real-time behaviour against timing requirements with tolerances. Our approach to specifying and verifying FBs~\cite{PangWangLawfordWassyng2013}, compared with others on verifying PLC programs in contexts of model checking and theorem proving, is novel in three aspects: (1) extent of the case study; (2) practical application in the safety-critical industry; and (3) mature tool support of theorem proving.

In our formal setting, proving that an FBD implementation is correct (with respect to its intended input-output timing requirements) is essentially proving that it is a valid refinement. However, our purpose of verification is on the observable input-output behaviour, as opposed other properties such as boundedness, liveness, and robustness (e.g., \cite{Ding2013, Wijs0213, shrinktech, so38269}). Of more relevance is the use of timed automata to model timing tolerances with ASAP (as soon as possible) semantics to verify the correctness of implementation~\cite{Wulf05}, but with no suggestion for either tool support or its adoption in practice.  

%% file: input_sec_conclusion.tex
\section{Conclusion}\label{sec:conclusion}

In this paper we report our application of a formal approach on using FBs (including timers) from IEC~61131-3 to verify two subsystems of an industrial software control system from the nuclear domain. We re-formalize all three IEC~61131-3 timers to incorporate the notion of tolerances. Specifically, we use the re-formalized IEC~61131 on-delay timer for the proposed FBD implementations, and prove that they are feasible and correct (i.e.,~satisfies the intended timing requirements). While attempting to verify the two subsystems, we find an issue of initialization failure, and an issue of missing implementation assumption. In both cases, we suggest possible solutions. We identify patterns of proof commands that are amenable to strategies that will facilitate the automated verification of the feasibility and correctness of other subsystems. As ongoing and future work, we first aim to verify subsystems with more sophisticated timing requirements, e.g., nested \name{Held\_For} expressions. Second, we aim to prove safety properties from the composition of real-time subsystems. Third, we aim to automate the process of proofs that share a common structure. 


%% file: ESSS_2015.bbl
\begin{thebibliography}{10}
\providecommand{\bibitemdeclare}[2]{}
\providecommand{\surnamestart}{}
\providecommand{\surnameend}{}
\providecommand{\urlprefix}{Available at }
\providecommand{\url}[1]{\texttt{#1}}
\providecommand{\href}[2]{\texttt{#2}}
\providecommand{\urlalt}[2]{\href{#1}{#2}}
\providecommand{\doi}[1]{doi:\urlalt{http://dx.doi.org/#1}{#1}}
\providecommand{\bibinfo}[2]{#2}

\bibitemdeclare{misc}{DO178C}
\bibitem{DO178C}
 (\bibinfo{year}{2011}): \emph{\bibinfo{title}{{DO-178C: Software
  Considerations in Airborne Systems and Equipment Certification}}}.
\newblock \bibinfo{howpublished}{Special Committee 205 of RTCA}.

\bibitemdeclare{article}{so38269}
\bibitem{so38269}
\bibinfo{author}{Ed~\surnamestart {Brinksma}\surnameend},
  \bibinfo{author}{Angelika \surnamestart {Mader}\surnameend} \&
  \bibinfo{author}{Ansgar \surnamestart {Fehnker}\surnameend}
  (\bibinfo{year}{2002}): \emph{\bibinfo{title}{Verification and optimization
  of a PLC control schedule}}.
\newblock {\sl \bibinfo{journal}{International Journal on Software Tools for
  Technology Transfer (STTT)}} \bibinfo{volume}{4}(\bibinfo{number}{1}), pp.
  \bibinfo{pages}{21--33}.
\newblock \urlprefix\url{http://dx.doi.org/10.1007/s10009-002-0079-0}.

\bibitemdeclare{article}{Ding2013}
\bibitem{Ding2013}
\bibinfo{author}{Zhijun \surnamestart Ding\surnameend},
  \bibinfo{author}{Changjun \surnamestart Jiang\surnameend} \&
  \bibinfo{author}{Mengchu \surnamestart Zhou\surnameend}
  (\bibinfo{year}{2013}): \emph{\bibinfo{title}{Design, Analysis and
  Verification of Real-Time Systems Based on Time Petri Net Refinement}}.
\newblock {\sl \bibinfo{journal}{ACM Trans. Embed. Comput. Syst.}}
  \bibinfo{volume}{12}(\bibinfo{number}{1}), pp. \bibinfo{pages}{4:1--4:18}.
\newblock \urlprefix\url{http://dx.doi.org/10.1145/2406336.2406340}.

\bibitemdeclare{phdthesis}{Hu08}
\bibitem{Hu08}
\bibinfo{author}{Xiayong \surnamestart Hu\surnameend} (\bibinfo{year}{2008}):
  \emph{\bibinfo{title}{Proving implementability of timing properties with
  tolerance}}.
\newblock Ph.D. thesis, \bibinfo{school}{McMaster University, Department of
  Computing and Software}.

\bibitemdeclare{inproceedings}{HuLawfordWassyng2008}
\bibitem{HuLawfordWassyng2008}
\bibinfo{author}{Xiayong \surnamestart Hu\surnameend}, \bibinfo{author}{Mark
  \surnamestart Lawford\surnameend} \& \bibinfo{author}{Alan \surnamestart
  Wassyng\surnameend} (\bibinfo{year}{2009}): \emph{\bibinfo{title}{Formal
  Verification of the Implementability of Timing Requirements}}.
\newblock In: {\sl \bibinfo{booktitle}{FMICS}}, {\sl \bibinfo{series}{{LNCS}}}
  \bibinfo{volume}{5596}, \bibinfo{publisher}{Springer}, pp.
  \bibinfo{pages}{119--134}.
\newblock \urlprefix\url{http://dx.doi.org/10.1007/978-3-642-03240-0\_12}.

\bibitemdeclare{book}{IEC:2003:IEP}
\bibitem{IEC:2003:IEP}
\bibinfo{author}{\surnamestart {IEC}\surnameend} (\bibinfo{year}{2003}):
  \emph{\bibinfo{title}{61131-3 Ed. 2.0 en:2003: Programmable Controllers ---
  Part 3: Programming Languages}}.
\newblock \bibinfo{publisher}{International Electrotechnical Commission}.

\bibitemdeclare{article}{Jin2010980}
\bibitem{Jin2010980}
\bibinfo{author}{Ying \surnamestart Jin\surnameend} \&
  \bibinfo{author}{David~Lorge \surnamestart Parnas\surnameend}
  (\bibinfo{year}{2010}): \emph{\bibinfo{title}{Defining The Meaning of Tabular
  Mathematical Expressions}}.
\newblock {\sl \bibinfo{journal}{Science of Computer Programming}}
  \bibinfo{volume}{75}(\bibinfo{number}{11}), pp. \bibinfo{pages}{980 -- 1000}.
\newblock \urlprefix\url{http://dx.doi.org/10.1016/j.scico.2009.12.009}.

\bibitemdeclare{inproceedings}{LMFM00}
\bibitem{LMFM00}
\bibinfo{author}{Mark \surnamestart Lawford\surnameend}, \bibinfo{author}{Jeff
  \surnamestart McDougall\surnameend}, \bibinfo{author}{Peter \surnamestart
  Froebel\surnameend} \& \bibinfo{author}{Greg \surnamestart Moum\surnameend}
  (\bibinfo{year}{2000}): \emph{\bibinfo{title}{Practical application of
  functional and relational methods for the specification and verification of
  safety critical software}}.
\newblock In: {\sl \bibinfo{booktitle}{Proc.\ of AMAST 2000}}, {\sl
  \bibinfo{series}{LNCS}} \bibinfo{volume}{1816},
  \bibinfo{publisher}{Springer}, pp. \bibinfo{pages}{73--88}.
\newblock \urlprefix\url{http://dx.doi.org/10.1007/3-540-45499-3\_8}.

\bibitemdeclare{inproceedings}{Owre1992}
\bibitem{Owre1992}
\bibinfo{author}{Sam \surnamestart Owre\surnameend}, \bibinfo{author}{John~M.
  \surnamestart Rushby\surnameend} \& \bibinfo{author}{Natarajan \surnamestart
  Shankar\surnameend} (\bibinfo{year}{1992}): \emph{\bibinfo{title}{{PVS: A
  Prototype Verification System}}}.
\newblock In: {\sl \bibinfo{booktitle}{CADE}}, {\sl \bibinfo{series}{LNCS}}
  \bibinfo{volume}{607}, pp. \bibinfo{pages}{748--752}.
\newblock \urlprefix\url{http://dx.doi.org/10.1007/3-540-55602-8\_217}.

\bibitemdeclare{inproceedings}{PangWangLawfordWassyng2013}
\bibitem{PangWangLawfordWassyng2013}
\bibinfo{author}{Linna \surnamestart Pang\surnameend},
  \bibinfo{author}{Chen-Wei \surnamestart Wang\surnameend},
  \bibinfo{author}{Mark \surnamestart Lawford\surnameend} \&
  \bibinfo{author}{Alan \surnamestart Wassyng\surnameend}
  (\bibinfo{year}{2013}): \emph{\bibinfo{title}{{Formalizing and Verifying
  Function Blocks using Tabular Expressions and PVS}}}.
\newblock In: {\sl \bibinfo{booktitle}{FTSCS}}, {\sl
  \bibinfo{series}{Communications in Computer and Information Science}}
  \bibinfo{volume}{419}, \bibinfo{publisher}{Spring}, pp.
  \bibinfo{pages}{163--178}.
\newblock \urlprefix\url{http://dx.doi.org/10.1007/978-3-319-05416-2\_9}.

\bibitemdeclare{techreport}{PangWangLawfordWassyngTechReport2014}
\bibitem{PangWangLawfordWassyngTechReport2014}
\bibinfo{author}{Linna \surnamestart Pang\surnameend},
  \bibinfo{author}{Chen-Wei \surnamestart Wang\surnameend},
  \bibinfo{author}{Mark \surnamestart Lawford\surnameend},
  \bibinfo{author}{Alan \surnamestart Wassyng\surnameend},
  \bibinfo{author}{Josh \surnamestart Newell\surnameend}, \bibinfo{author}{Vera
  \surnamestart Chow\surnameend} \& \bibinfo{author}{David \surnamestart
  Tremaine\surnameend} (\bibinfo{year}{2014}): \emph{\bibinfo{title}{{Formal
  Verification of Real-Time Function Blocks using PVS}}}.
\newblock \bibinfo{type}{Technical Report} \bibinfo{number}{16},
  \bibinfo{institution}{McSCert}.
\newblock
  \bibinfo{note}{\url{https://www.mcscert.ca/index.php/documents/mcscert-repor%
ts?view=publication&task=show&id=16}}.

\bibitemdeclare{article}{Parnas:1994:PDW:203102.203107}
\bibitem{Parnas:1994:PDW:203102.203107}
\bibinfo{author}{David~Lorge \surnamestart Parnas\surnameend},
  \bibinfo{author}{Jan \surnamestart Madey\surnameend} \&
  \bibinfo{author}{Michal \surnamestart Iglewski\surnameend}
  (\bibinfo{year}{1994}): \emph{\bibinfo{title}{Precise Documentation of
  Well-Structured Programs}}.
\newblock {\sl \bibinfo{journal}{IEEE Transactions on Software Engineering}}
  \bibinfo{volume}{20}, pp. \bibinfo{pages}{948--976}.
\newblock \urlprefix\url{http://dx.doi.org/10.1109/32.368133}.

\bibitemdeclare{incollection}{shrinktech}
\bibitem{shrinktech}
\bibinfo{author}{Ocan \surnamestart Sankur\surnameend} (\bibinfo{year}{2013}):
  \emph{\bibinfo{title}{Shrinktech: A Tool for the Robustness Analysis of Timed
  Automata}}.
\newblock In: {\sl \bibinfo{booktitle}{Computer Aided Verification}}, {\sl
  \bibinfo{series}{{LNCS}}} \bibinfo{volume}{8044},
  \bibinfo{publisher}{Springer}, pp. \bibinfo{pages}{1006--1012}.
\newblock \urlprefix\url{http://dx.doi.org/10.1007/978-3-642-39799-8\_72}.

\bibitemdeclare{inproceedings}{WasLaw:2003}
\bibitem{WasLaw:2003}
\bibinfo{author}{Alan \surnamestart Wassyng\surnameend} \&
  \bibinfo{author}{Mark \surnamestart Lawford\surnameend}
  (\bibinfo{year}{2003}): \emph{\bibinfo{title}{Lessons Learned from a
  Successful Implementation of Formal Methods in an Industrial Project}}.
\newblock In: {\sl \bibinfo{booktitle}{FME 2003}}, {\sl \bibinfo{series}{LNCS}}
  \bibinfo{volume}{2805}, \bibinfo{publisher}{Springer}, pp.
  \bibinfo{pages}{133--153}.
\newblock \urlprefix\url{http://dx.doi.org/10.1007/978-3-540-45236-2\_9}.

\bibitemdeclare{inproceedings}{Wassyng05}
\bibitem{Wassyng05}
\bibinfo{author}{Alan \surnamestart Wassyng\surnameend}, \bibinfo{author}{Mark
  \surnamestart Lawford\surnameend} \& \bibinfo{author}{Xiaoyong \surnamestart
  Hu\surnameend} (\bibinfo{year}{2005}): \emph{\bibinfo{title}{Timing
  Tolerances in Safety-Critical Software}}.
\newblock In: {\sl \bibinfo{booktitle}{FM 2005}}, {\sl \bibinfo{series}{LNCS}}
  \bibinfo{volume}{3582}, \bibinfo{publisher}{Springer}, pp.
  \bibinfo{pages}{157 -- 172}.
\newblock \urlprefix\url{http://dx.doi.org/10.1007/11526841\_12}.

\bibitemdeclare{incollection}{Wijs0213}
\bibitem{Wijs0213}
\bibinfo{author}{Anton \surnamestart Wijs\surnameend} \& \bibinfo{author}{Luc
  \surnamestart Engelen\surnameend} (\bibinfo{year}{2013}):
  \emph{\bibinfo{title}{Efficient Property Preservation Checking of Model
  Refinements}}.
\newblock In: {\sl \bibinfo{booktitle}{TACAS}}, {\sl \bibinfo{series}{LNCS}}
  \bibinfo{volume}{7795}, \bibinfo{publisher}{Springer}, pp.
  \bibinfo{pages}{565--579}.
\newblock \urlprefix\url{http://dx.doi.org/10.1007/978-3-642-36742-7\_41}.

\bibitemdeclare{article}{Wulf05}
\bibitem{Wulf05}
\bibinfo{author}{Martin~De \surnamestart Wulf\surnameend},
  \bibinfo{author}{Laurent \surnamestart Doyen\surnameend} \&
  \bibinfo{author}{Jean-François \surnamestart Raskin\surnameend}
  (\bibinfo{year}{2005}): \emph{\bibinfo{title}{Almost {ASAP} semantics: from
  timed models to timed implementations}}.
\newblock {\sl \bibinfo{journal}{FAC}}
  \bibinfo{volume}{17}(\bibinfo{number}{3}), pp. \bibinfo{pages}{319--341}.
\newblock \urlprefix\url{http://dx.doi.org/10.1007/978-3-540-24743-2\_20}.

\end{thebibliography}
